# Appsent: A Tool That Analyzes App Reviews

Saurabh Malgaonkar, Chan Won Lee, Sherlock A. Licorish, Bastin Tony Roy Savarimuthu
Department of Information Science
University of Otago, New Zealand
saurabh.malgaonkar@postgrad.otago.ac.nz,
lchacha04@hotmail.com,
{sherlock.licorish,tony.savarimuthu}@otago.ac.nz

Amjed Tahir
School of Engineering and Advanced Technology
Massey University, New Zealand
a.tahir@massey.ac.nz

*Abstract*—Enterprises are always on the lookout for tools that analyze end-users' perspectives on their products. In particular, app reviews have been assessed as useful for guiding improvement efforts and software evolution, however, developers find reading app reviews to be a labor intensive exercise. If such a barrier is eliminated, however, evidence shows that responding to reviews enhances end-users' satisfaction and contributes towards the success of products. In this paper, we present *Appsent*, a mobile analytics tool (as an app), to facilitate the analysis of app reviews. This development was led by a literature review on the problem and subsequent evaluation of current available solutions to this challenge. Our investigation found that there was scope to extend currently available tools that analyze app reviews. These gaps thus informed the design and development of Appsent. We subsequently performed an empirical evaluation to validate Appsent's usability and the helpfulness of analytics features from users' perspective. Outcomes of this evaluation reveal that Appsent provides user-friendly interfaces, helpful functionalities and meaningful analytics. Appsent extracts and visualizes important perceptions from end-users' feedback, identifying insights into end-users' opinions about various aspects of software features. Although Appsent was developed as a prototype for analyzing app reviews, this tool may be of utility for analyzing product reviews more generally.

## I. INTRODUCTION

The mobile app market was estimated to be a 77 billion dollar industry in 2017, with more than five million apps hosted on Online Application Distribution Platforms (OADPs) [1, 2]. This is linked to massive sales of mobile devices and popularity of their usage worldwide [3]. The commonly accessed OADPs are Google play store, Apple app store and Windows apps [4-6]. OADPs allow developers to host their apps for mobile device users, facilitating app updates when developers release new versions of their apps. Another useful feature provided by OADPs is the ability to support direct communication between app users and developers through reviews. These reviews contain important feedback related to apps' performance from users' perspectives. However, OADPs host numerous reviews, which are open to public access in informing future users' decisions in relation to app use. Beyond a star rating, reviews normally contain complaints about commonly faced problems, users' sentiments about features, suggestions for improvement, and requests for new features [4].

Thus, in meeting the expectations of users, app developers must analyze users' reviews to evolve their apps. This knowledge also significantly assists developers in their user-driven software quality evaluation and product marketing process [5]. However, manually processing large volumes of reviews demands high levels of cognitive load and time if it is performed manually [6, 7]. In fact, this burden may also be compounded due to ambiguity and sarcasm present in the reviews [8, 9].

Thus, a combination of natural language processing and data mining techniques have been recommended for addressing such challenges [9-11]. Such techniques are held to have promise in terms of assisting developers with extracting and visualizing meaningful information from user reviews [12]. This way, developers may quickly identify issues faced by users and discern the source of their (dis)satisfaction, in enhancing the quality of their app and gaining a competitive edge in the app market [13, 14]. We have looked to validate this opportunity in this work, and developed Appsent to provide these details for developers. We introduce Appsent in this paper, whose primary objective is to provide instant real-time meaningful interpretation of the information present in the app reviews. Although being a prototype, outcomes of Appsent's evaluations suggest that this tool could be of utility to app developers. We present our portfolio of work around the development of Appsent and its evaluation, contributing practical insights for developers, and understandings for the software engineering community in terms of how tools may be engineered to rapidly support the evolution of software.

The remaining sections of this paper are organized as follows. In Section II, we review relevant literature and tools for analyzing user reviews. This leads to the identification of research gaps and suitable research questions. Section III describes the design and implementation aspects of Appsent. Section IV presents our empirical evaluation of Appsent, followed by the tool's overview and evaluation outcomes in Section V. We discuss our findings and implications of the work in Section VI, before considering threats to the work in Section VII. Finally, we provide concluding remarks in Section VIII.

## II. BACKGROUND AND RESEARCH QUESTIONS

Text mining as a discipline is held to offer an efficient solution to the problem of processing large volumes of textual information to extract knowledge [15]. The text mining process is data-driven, and assists significantly in identifying the hidden patterns or trends in unstructured text [16]. Central to the meaningful interpretation of textual information is the ability to identify the polarity of users. The primary objective of polarity analysis is to discover the attitude of the user towards specific things or events [17]. Polarity can be evaluated via sentiment (e.g., positive or negative) or emotion (e.g., happy or fear). The polarity of a sample of text may be uncovered by using learning- or lexicon-based methods [18]. Learning methods build a feature classification model by utilizing machine learning techniques or probabilistic models, whereas lexicon methods use well-built dictionaries to determine the meaning of textual entities.

Another critical stage of text mining is topic or feature identification and extraction. Part-of-speech (POS) tagging and n-gram analysis are commonly used for extracting features or topics of interest in textual data [19]. For instance, Licorish et al. [6] employed these approaches to identify the most frequently mentioned features in enhancement requests logged by the Android community. In another study, Iacob et al. [19] were successful in discovering associations among features through the use of these approaches and Latent Dirichlet Allocation (LDA) models. A more advanced study by Lee et al. [20] has deployed POS tagging, n-gram analysis and Social Network Analysis (SNA) techniques to identify features and explore associations among features that were mentioned in reviews. This latter study places emphasis on the importance of discovering the relations between features in understanding the complexities of defects and improvement pointers that are identified by users in reviews. The abovementioned studies automatically process reviews to provide a high-level view of the information of interest to users. Beyond understanding features that capture users' interests, such provisions enable developers to identify the complex relationships between app features, and significantly assists in co-relating users' problems. We anticipated that tools built to assist developers with extracting and visualizing meaningful information from user reviews should identify users' opinions about software features (e.g., positive views around specific features), identify the features of particular interest (e.g., broken features), and discover the associations among the features (e.g., which group of features affects each other).

Browsing for such products on the app stores and wider internet it is observed that software enterprises have used the methods identified above to develop tools that fulfill the task of analyzing the performance of apps released on OADPs. After identifying four popular tools available online, we performed a systematic comparative study of these tools. The four tools are AppTrace, AppFigures, SensorTower, and Apptentive [21-24]. Table I highlights the systematic comparative study of the four tools. The tools have been classified based on their target audience, market pricing, domain of application, and a set of features. We compared the features organically, and also used dimensions suggested in [20] (noted below) for evaluating the tools utility for identifying users' opinions about software features, features of particular interest, and associations among features. We compared the following criteria:

- *daily ranks*: computed based on the number of daily downloads.
- *country ranks*: computed based on the number of downloads across countries.
- *sentiment analysis and sentiment analysis with keywords*: computed based on the polarity of reviews and the polarity that keywords attracted.
- *sales analysis*: computed based on revenue generated by the app to date.
- *competitor analysis*: computed based on the ranking, sentiment, and sales features.
- *star ratings*: the rating of the app on a numerical scale of one to five.
- *advanced text mining option*: the ability of the tool to discover meaningful associations among the entities of interests present in app reviews.

We also informally evaluated if the tool was provided as an interface for mobile devices, given their popularity [1, 2]. In Table I, '0' denotes the absence of a feature while '1' indicates the presence of a feature in the particular tool.

Table I. Summary of commercially available tools to analyze reviews

| | | 1 = Feature Present, 0 = Feature Absent | | | |
|---|---|---|---|---|---|
| **Application Name** | | App Trace | App Figures | Sensor Tower | Apptentive |
| **Target Audience** | | Product owner, Developer | Product owner, Developer | Product owner, Developer | Product owner, Developer |
| **Price** | | Free | $9/month | $79/month (Personal) $399/month (Business) | $299/month |
| **Domain** | | Smart phone Apps | Smart phone Apps | Smart phone Apps | Smart phone Apps |
| FEATURES | Daily Ranks | 1 | 1 | 1 | 1 |
| | Country Ranks | 1 | 1 | 1 | 1 |
| | Sentiment Analysis | 1 | 1 | 1 | 0 |
| | Sentiment Analysis with Keywords | 1 | 0 | 0 | 0 |
| | Sales Analysis | 0 | 1 | 1 | 0 |
| | Competitor Analysis | 0 | 0 | 1 | 0 |
| | Star Rating | 0 | 0 | 0 | 1 |
| | Advanced Text Mining (Feature relationship analysis) | 0 | 0 | 0 | 0 |

Evaluating the tools it was observed that only Apptentive was available for installation and use on smartphone (mobile) devices (Table I). However, overall, Apptentive supported fewer features than the other tools. By running the tools, we found that the sentiment analysis feature of AppTrace did not return readable visualizations at times (i.e., too much dimensions were evident). AppFigures required manual intervention for analyzing reviews, and SensorTower could not fully distinguish the sentiment for some reviews, requiring manual intervention. Overall, it was observed that these tools are still being evolved and they do not provide mechanisms for identifying associations among features, which is assessed as critical for identifying complex relationships among defects [20].

Therefore, we believe that having a tool that analyzes user reviews will allow developers to quickly identify issues faced by users and discern the source of their (dis)satisfaction, in enhancing the quality of their app and gaining a competitive edge in the apps market. However, examining previous works and tools available for analyzing reviews to inform developers, Table I shows that while tools available provide various functions to support reviews analysis, they lack various features that may be deemed pertinent to developers [20]. We thus proposed to provide a prototype app to bridge this gap. To guide this effort we formulated the following three research questions:

*RQ1. How can we design and develop a mobile app for analyzing reviews that is deemed highly usable?*

Upon the creation of the mobile app, we investigate the utility provided by the functionalities offered by the app. Thus, the second question we investigated is:

*RQ2. Does the newly developed app provide functionalities that are helpful?*

Any analytics tool that is developed should unearth insights from data and present it to the users. In this work, we investigate the ability of the tool to provide insights at different levels (e.g., based on sentiments and emotions expressed by users). Towards examining the nature of insights provided by the app, we pose the last question.

*RQ3. Does the developed app provide suitable interpretative analytics on multiple levels?*

### III. APPSENT DESIGN AND IMPLEMENTATION

We provide details around the techniques and principles that were used to design and implement our app, Appsent, in this section towards answering RQ1.

#### A. Appsent Design

App reviews contain unstructured textual information that needs to be pre-processed for relevance [25]. This process involves the removal of blank spaces, unwanted characters, and stopwords. We performed this process, before aggregating similar words (i.e., stemming). Stopwords were identified through WordNet and Stanford NLP API was used for stemming based on Porter's algorithm [26, 27]. This process refined reviews making them suitable for further analysis. We next performed natural language processing through the use of POS tagging [28]. For this phase we used the Stanford API to extract the nouns and verbs in reviews; which map to features and issues respectively [29]. The extracted features and issues were tallied to quantify the frequency of each feature or issue. Sentiment analysis identifies the positive, negative or neutral tone of a statement made by users. To detect the sentiments in reviews we employed a sentiment analysis method that has been inspired by the sentiment treebank technique [30]. In exploring emotions in reviews we used WordNet's lexicon-based method to identify six emotions, namely *happy*, *sad*, *surprise*, *fear*, *anger* and *disgust* [31]. In checking the associations among features we performed co-occurrence analysis as proposed by Lee et al. [20]. This analysis captures the relationship between features and associated problems through feature-feature and feature-issue analyses. Appsent implements the same strategy and performs co-occurrence analysis by considering the *sentiment*, *emotion*, and the *rating* (1-5) factors.

After finalizing the functional aspects of Appsent, it was necessary to design a user-friendly graphical user interface (GUI). We considered both users' interaction with the app and the provision of meaningful visualizations of results on mobile device screens. In informing our design we adapted Shneiderman's rules [32]. These rules are provided in Table II, covering the creation of shortcuts of the most frequently visited features of the app to designing the app for enjoyment. Of particular interest to us was also the provision of informative responses, memory optimization, eliminating unwanted details, quick load of the application, and the provision of highly immersive and appealing interfaces (refer to Table II).

Table II. Appsent's User Interface Design Rules

| ID. | Rule |
|---|---|
| 1 | Enable frequent users to use shortcuts |
| 2 | Offer informative feedback |
| 3 | Design dialogs to yield closure |
| 4 | Support internal locus of control |
| 5 | Consistency |
| 6 | Reversal of actions |
| 7 | Error prevention and simple error handling |
| 8 | Reduce short-term memory load |
| 9 | Design for multiple and dynamic contexts |
| 10 | Design for small devices |
| 11 | Design for limited and split attention |
| 12 | Design for speed and recovery |
| 13 | Design for "top-down" interaction |
| 14 | Allow for personalization |
| 15 | Design for enjoyment |

Figure 1 shows the diagrammatic representation of Appsent's system model, and highlights the various steps involved in the processing of app reviews. As illustrated in Figure 1, the unstructured textual data from reviews are pre-processed. This leads to a reduction in the size of data and removal of irrelevant data. After the data has been pre-processed, the data is then split into sentences, where each sentence is analyzed for emotional and sentiment states. Sentences present in reviews are parsed independently to identify the parts-of-speech (e.g., nouns and verbs). The POS tagger then marks the words (features or issues) of interests (e.g., "battery" and "drain"), and these words are stored in a database for further analysis. Thereafter, appropriate analysis mechanisms are executed to generate the information of interest. This information is stored in the smartphone's database. Appsent then queries the database as required, and the retrieved results are visualized using the appropriate data analytics charts. Thus, the smartphone app (Appsent) acts as a mobile carrier of the meaningful analytics information.

*B. Appsent Implementation*

Appsent was developed using Android studio and several open-source libraries [33]. To store data we used the SQLite database [34]. We employed two libraries for natural language processing; for POS tagging and sentiment extraction, we used Stanford CoreNLP 3.6.0 [35], and for performing emotion analysis we used Synesketch 2.0 [36]. For data visualization we used libraries from MPAndroidChart [37]. All the libraries used in this project are free for academic use. Appsent is built to operate on mobile devices running on Android version 3.2 (Honeycomb) or later, with a minimum of 1GB of memory.

## IV. APPSENT EMPIRICAL EVALUATION SETUP

In order to evaluate Appsent we engaged a company that designs mobile games in Dunedin, New Zealand called Runaway [38]. The company provided us with a dataset containing reviews taken from *Google Play* and Apple's *App Store* for one of their products. The product in question allows users to play a game involving nature. Our dataset contained 52,705 reviews with fifteen attributes associated with each review, including: Package Name, App Version Code, Reviewer Language, Device, Review Date, Star Rating, Review Title, and Review Text. We selected the *Review Date*, *Star Rating*, and *Review Text* attributes for our evaluation of our prototype (Appsent). *Review Date* and *Star Rating* contained values that were well structured (e.g., "July 5, 2018" and "4"), and did not require any pre-processing. However, *Review Text* is mostly subjected to pre-preprocessing as mentioned in Section III(A). Beyond pre-processing, the processes highlighted in Figure 1 were followed towards visualizing analyzed reviews on Appsent's display screens.

To gain quality feedback on Appsent's usability, helpfulness, and analytics features, we designed multiple questionnaires. The questionnaires were composed based on the guidelines provided by Ghazanfari et al. [39], and evaluated to answer our research questions in Section II. Questions in the questionnaires were designed using Likert scale measures comprising 1 to 5, where 1 stands for strong disagreement, and 5 stands for strong agreement. The questionnaires were categorized into three sections as shown in Tables III, IV and V, respectively. Having used the questionnaire to gather data to evaluate Appsent, we performed statistical analysis to evaluate the tool's utility based on the information gathered from the questionnaire [40].

Table III. Appsent's Usability Questionnaire

| ID | Question |
|---|---|
| Q1 | I think that I would like to use this system frequently. |
| Q2 | I thought the system was easy to use. |
| Q3 | I think that I would need the support of a technical person to be able to use this system. |
| Q4 | I found the system very cumbersome to use. |
| Q5 | I felt very confident using the system. |
| Q6 | I needed to learn a lot of things before I could get going with this system. |

Table IV. Appsent's Features' Helpfulness Questionnaire

| ID | Question |
|---|---|
| Q7 | The system provides me with multiple levels of information (fine-grained and aggregate information) about users' reviews. |
| Q8 | Using the system to analyze users' review requires less cognitive load for interpretation and is easy to understand than analyzing natural language text (i.e., reviews) manually. |
| Q9 | The visualizations that are used to represent results from review analytics (e.g., Bar graph, Line graph, and Pie graph) are appropriate and easy to understand. |
| Q10 | The system provides information in a timely fashion. |

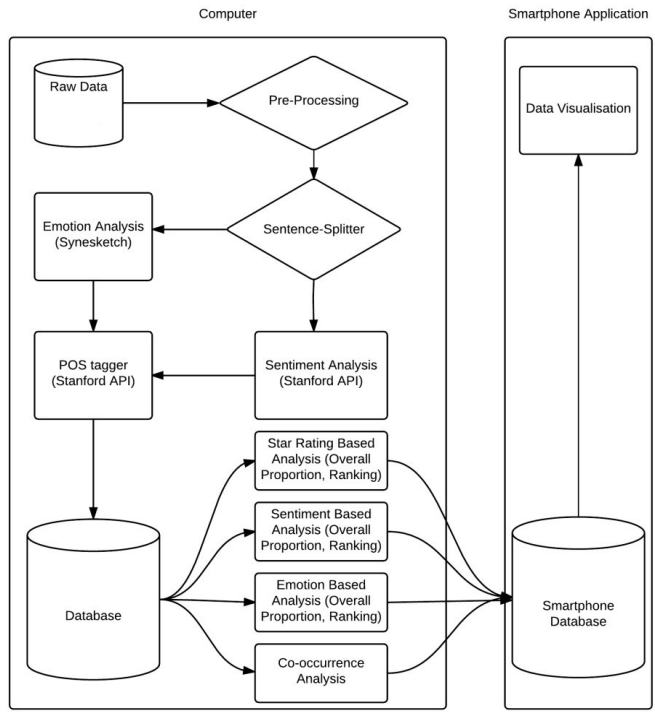

Figure 1. Appsent's system model

Table V. Appsent's Advanced Analytics Feature Evaluation Questionnaire

| ID | Question |
|---|---|
| Q11 | Studying the fine-grained sentiments (positive, negative and neutral) for the top features helped me to understand features that are received well and those that need improvements. |
| Q12 | Studying the fine-grained emotions (happy, sad, anger, fear, disgust and surprise) for top features helped me to understand features that are received well and those that need improvements. |
| Q13 | Timeline analysis on users' emotion, sentiment and star rating help me to understand the changes in users' sentiments and emotions over time. |
| Q14 | Studying the relationships (co-occurring terms) between features and associated issues helps with understanding interconnections between features and issues. |
| Q15 | Overall, the features provided by the system help me to identify improvements to be made to my app/products in order to satisfy users' need. |

We advertised for potential participants (app developers) for evaluating Appsent, and 15 respondents agreed to evaluate the tool. The participants were all aware and had previously interfaced with review portals, mobile devices and apps. In addition, all of the participants used mobile devices daily, and were required to report on their prior experience analyzing customer reviews. This allowed us to assess the feedback of those with experience (9 participants) versus those that possessed less knowledge (6 participants). Before performing the evaluation participants watched a brief video tutorial that introduced the basic functionalities of Appsent. This video tutorial acted as a user-manual for participants, and is available online[1]. For the actual evaluation, all 15 participants were asked to use Appsent to analyze 52,705 app reviews (provided by the company above). Appsent was installed on a LG Nexus 5 smartphone with 2GB RAM, a resolution of 1080x1920 pixels and running on Android version 4.4.4 (KitKat). After participants had finished using Appsent, they were asked to answer the three questionnaires above (refer Tables III, IV and V). The scores submitted by each participant were then recorded for analysis.

## V. TOOL OVERVIEW AND EVALUATION OUTCOMES

In this section, we overview Appsent's interfaces and main functionalities in more detailed, following by an evaluation of the tool.

### A. Appsent Interfaces

Overall, Appsent provides five major functionalities: 1) overview summary for apps, 2) star ranking analysis, 3) sentiment ranking analysis, 4) emotion analysis, and 5) a tutorial feature. These features are included on the home screen of the app. We have not included the home screen Figure given space limitation, however, the video link below[1] provides further details. The left interface in Figure 2 highlights the results of sentiment analysis produced by Appsent. For the sample of reviews above (i.e., 52,705 reviews), 34.3% were classified as positive, 35.5% were classified as neutral, and 30.2% were classified as negative. The sentiments are also visualized based on a timeline scale on the right interface in Figure 2.

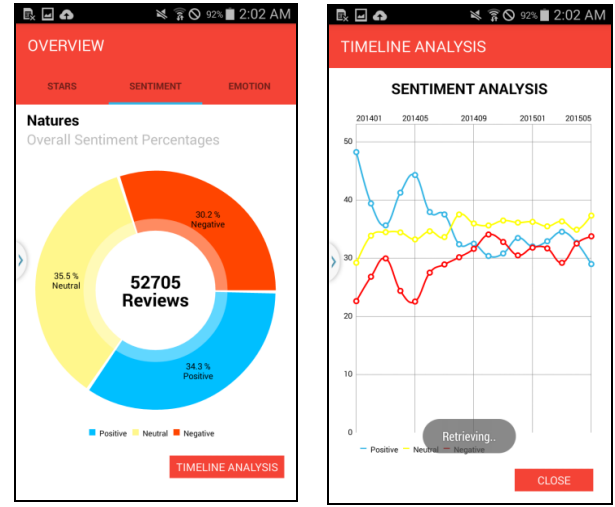

Figure 2. Appsent's sentiment analysis

The left interface in Figure 3 highlights the results of star ratings, with tabs across the top of the screen for each star rating (1 star to 5 star rating). Below the interface shows the top 10 features that attracted the specific star rating. For example, in Figure 3 it is shown that a specific feature attracted 1-star rating 182 times. Similar to the interface on the right in Figure 2 for sentiment analysis, the star ratings may also be analyzed on a timeline scale (refer to the right interface in Figure 3).

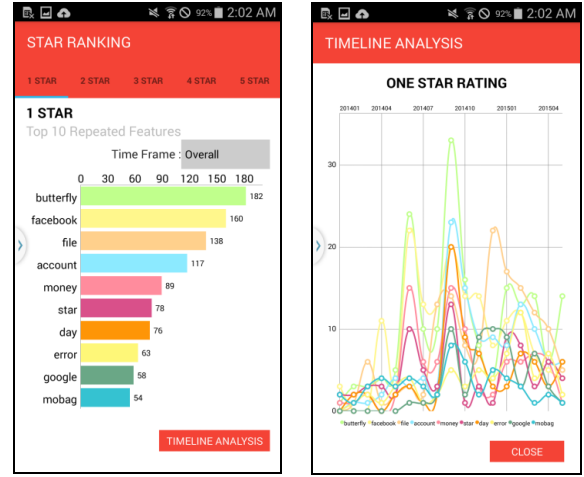

Figure 3. Appsent's star rating analysis

By clicking on the negative section (in red) of the pie chart in Figure 2, a different view is then shown (left interface in Figure 4). Here the top 10 negatively-rated features are presented (as for stars in Figure 3). However, when the *co-occurring* button is clicked, the defects or issues associated with the particular feature of interest are retrieved and displayed (refer to the right interface in Figure 4). So for instance, in Figure 4 the top defects associated with the feature butterfly are displayed.

---
[1] https://youtu.be/FPI3sBg0xBY

There is also an option provided on the Appsent's screen for the end-user to select a specific feature for further analysis in the form of a drop down list.

The co-occurring analysis comprises feature-issue (or defect) and feature-feature relationships. For instance, if "facebook" in Figure 4 is selected, the top defects related to this feature will then be shown in the app. In the screen evident on the right in Figure 4 butterfly was associated with "play", "spend", "update", "keep", and so on, as issues. Selecting "Features" at the top of the right interface in Figure 4 will also allow a user to visualize all of the features that were connected to the specific feature in question.

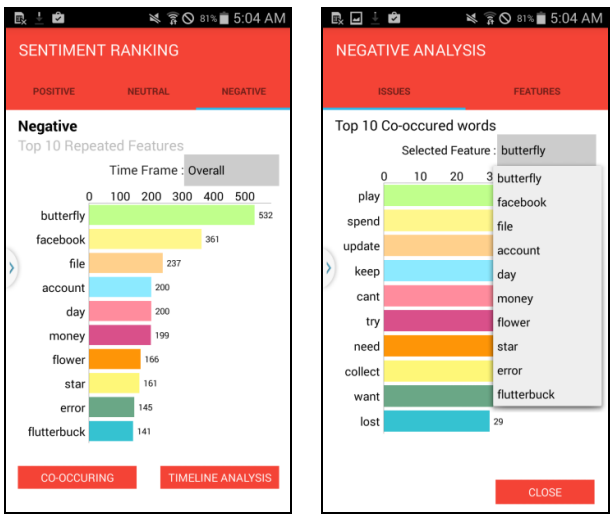

Figure 4. Appsent's sentiment and co-occurrence analysis

Users may go beyond sentiments and examine the correlation between sentiments and emotions for specific features (*sentiment* and *emotion* ranking screens are shown in Figure 5). For example, in Figure 5, 819 reviews expressed positive sentiments about the butterfly feature (on the left), while 893 demonstrated users' happiness (on the right).

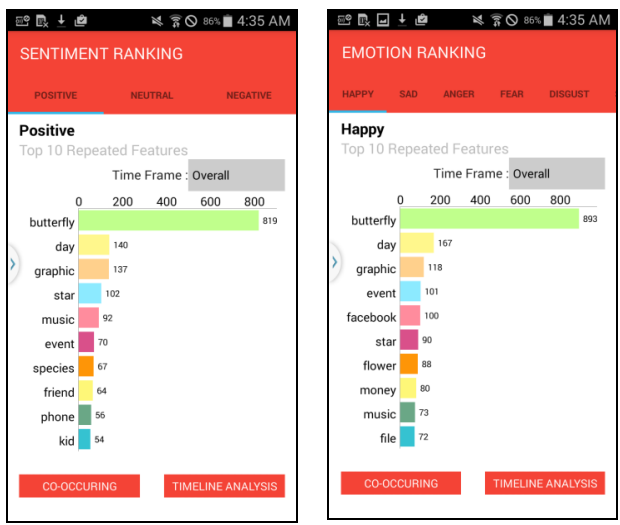

Figure 5. Appsent's sentiment-emotion (multidimensional) analysis

## B. Appsent Evaluation

We provide the evaluation outcomes for Appsent in this section. The evaluation covers the following three aspects: 1) usability, 2) helpfulness and 3) analytics features.

### 1) Appsent's usability scores

To evaluate participants' perception of Appsent's usability we computed the scores returned from the completed system usability scale (SUS) questionnaire in Table III. Here, six questions were asked, with respondent scores summed for each question. We stored the scores received from 0 to 4, matching the 1 to 5 Likert scales value selected before converting these to an overall percentage value. This approach was also followed in the analysis conducted in the remaining subsections below. Figure 6 provides a summary of Appsent's usability scores obtained from the experienced (Exp.), non-experienced (Non-Exp.), and overall participants (refer to Section IV for details about participants). Out of six questions, for overall participants, question 4 showed the highest score (Mean = 3.4, Median = 4.0, Standard Deviation = 1.0). The overall lowest score was seen for question 2 (Mean = 2.6, Median = 3.0, Standard Deviation = 0.9). However, for the experienced group, the lowest score was seen for question 1 (Mean= 2.4, Median = 2.0, Standard Deviation = 1.1).

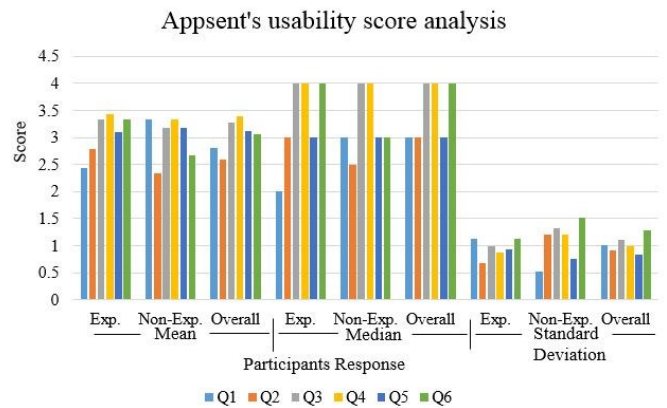

Figure 6. Appsent's usability score analysis

Figure 7 provides the overall usability scores for Appsent (usability assessment questions are shown in Table III). Appsent scored 76.1 for usability (Median = 79.2, Standard Deviation = 17.5). The group-based analysis indicates that the average score of participants that are experienced (Mean = 76.9) was slightly higher than for those who were non-experienced (Mean = 75.0). However, we have observed the opposite result in the median, where the median of participants with no experience (Median = 81.3) showed higher score than the participants with experience (Median = 75.0).

### 2) Appsent's helpfulness scores

To evaluate the helpfulness of the analytics features which are provided by Appsent, we asked participants to answer four questions (refer Table IV). Similar to the usability assessment, we have carried out analyses based on the experienced (Exp.) and non-experienced (Non-Exp.) categories of participants. Figure 8 visualizes these results, demonstrating that question 3 recorded the highest score (Mean = 3.5, Median = 4.0,

Standard Deviation = 0.8). The lowest score was seen for questions 1, 2 and 4 in which the values were equal (Mean = 3.4, Median = 4.0, Standard Deviation = 0.7~0.8). Overall, the score was very close to the maximum score (max = 4). From the group based analysis, we observed that the experienced participants' scores for all questions were higher than that of the non-experienced participants. The highest score in the experienced group is for question 2 (Mean = 3.8, Median = 4.0, Standard Deviation = 0.4), however, in the non-experienced group, the score for question 2 was the lowest (Mean = 2.8, Median = 3.0, Standard Deviation = 1.2). We have also observed a reversed pattern for the scores for question 1.

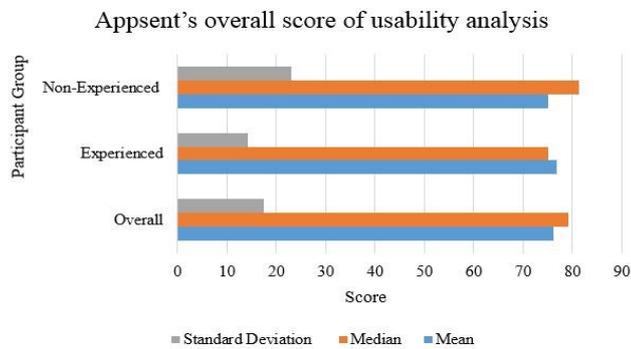

Figure 7. Appsent's overall score of usability analysis

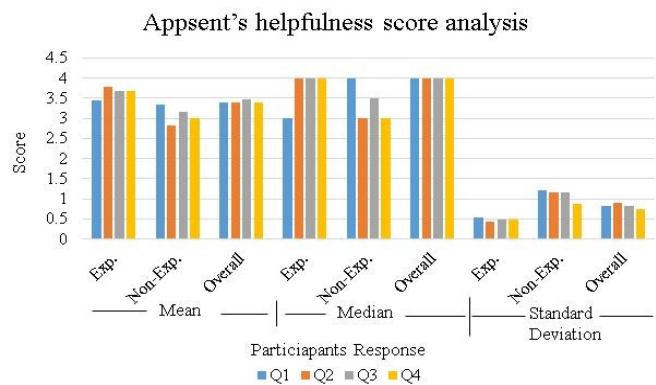

Figure 8. Appsent's helpfulness score analysis

Figure 9 presents Appsent's overall helpfulness scores, with an average of 85.4 returned by respondents (Median = 93.8, Standard Deviation = 17.1). The group-based measures indicate that experienced participants (Mean = 91.0, Median = 93.8, Standard Deviation = 9.4) ranked Appsent higher than those that were less experienced (Mean = 77.1, Median = 81.3, Standard Deviation = 23.3).

*3) Appsent's analytics scores*

To evaluate the analytics features of Appsent, we developed five questions in Section IV (refer Table V). Figure 10 provides a summary of participants' scores in response to these questions. Overall, out of the five questions, questions 3 and 4 show the highest average score (Q3: Mean = 3.5, Median = 4.0, Standard Deviation = 0.7; Q4: Mean = 3.5, Median = 4.0, Standard Deviation = 1.1). Questions 1 and 2 showed the lowest score (Mean = 3.1, Median = 3.0, Standard Deviation = 0.7~0.8). However, when comparing the means these scores are not remarkably different (the mean difference is 0.3). The experienced group (Exp.) assigned the highest score for question 4 (Mean = 3.9, Median = 4.0, Standard Deviation = 0.3) and the lowest score for question 2 (Mean = 3.3, Median = 3.0, Standard Deviation = 0.7). The non-experienced (Non-Exp.) group assigned the highest score for question 3 (Mean = 3.0, Median = 3.0, Standard = 0.9) and lowest score for question 1 (Mean = 2.5, Median = 3.0, Standard Deviation = 0.8).

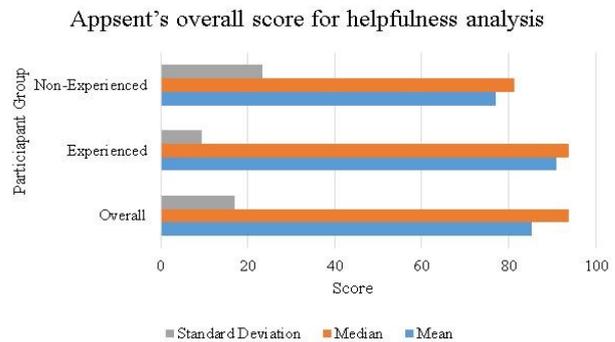

Figure 9. Appsent's overall score for helpfulness analysis

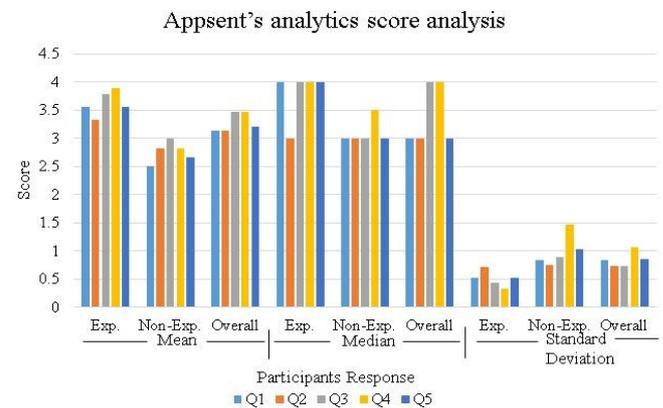

Figure 10. Appsent's analytics score analysis

Figure 11 provides the overall analytics score of Appsent, where it is shown that Appsent scored 82.0 on average for its analytics (Median = 85.0, Standard Deviation = 17.6). In addition, the median score returned for participants with experience (Median = 93.8) was higher than the score for participants with no experience (Median = 81.3).

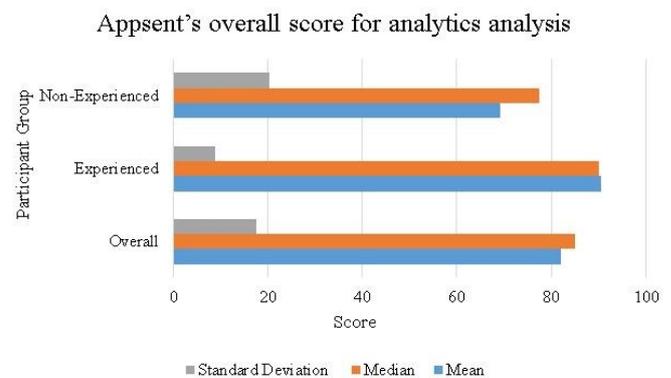

Figure 11. Appsent's overall score for analytics analysis

*4) Appsent's overall scores*

Figure 12 presents a comparison of scores that reflected Appsent's *usability*, *helpfulness* and *analytics* features. The scores of all the questionnaires were aggregated and then visualized. As shown in Figure 12, it is noted that Appsent's *helpfulness* was rated highest, with the analytics also being scored well by those that evaluated the app. Furthermore, while usability measures were the lowest, these score were also acceptable. Outcomes here suggest that Appsent could be of utility to those handling software reviews logged on app stores. We consider this issue at length in the following section.

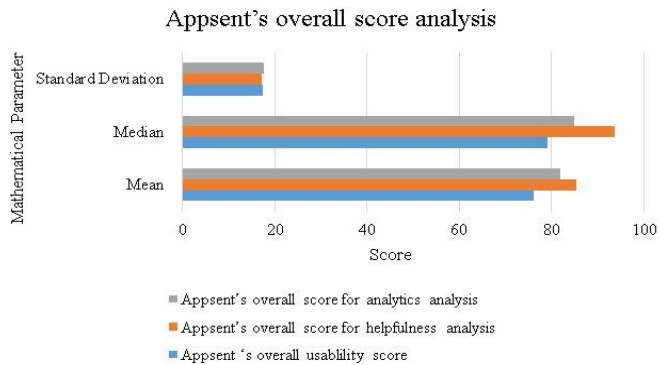

Figure 12. Appsent's overall score analysis

## VI. DISCUSSION AND IMPLICATIONS

We discuss our findings in answering our research questions in this section. Firstly, we discuss RQ1 where we examine Appsent's *usability* and its utility for analyzing reviews. We next discuss RQ2 and the *helpfulness* of Appsent. Finally, discussions for RQ3 are provided, where we review the actual *analytics* support that Appsent provides.

*RQ1. How can we design and develop a mobile app for analyzing reviews that is deemed highly usable?*

Results in the previous section show that the average score returned for Appsent's usability from the participants' evaluations was 76.1 (out of 100). This score shows that although Appsent is only developed as a prototype, it perceived to have good usability. That said, feedback also points to avenues for improving the tool. Overall, users that were aware of other analytics tools did not score Appsent to be much more usable than those that had encountered such a tool for the first time. In fact, the most substantial difference was observed in question one that asked: "*I think that I would like to use this system frequently*". For this particular question the experienced group had a lower score (2.4 out of 4) than non-experienced group (3.3 out of 4). We received additional feedback from the experienced group indicating that there was no requirement for real-time monitoring of users' opinion from the app-reviews, and hence they believed that there would be no need to use the tool frequently. In fact, the experienced group believed that Appsent is a good tool that they would be happy to use for analyzing bulk reviews periodically. The good feedback received for Appsent was perhaps directly related to the interface design rules that were used during its development (refer Table II). We deliberately aimed to reduce end-user actions and provided end-users with the information that was best suited to their needs [39]. We also ensured the provision of informative responses, memory optimization, eliminated unwanted details, quick loading of the application, and the provision of highly immersive and appealing screens. These principles ensured that Appsent was assessed as usable and the tool was well received.

On the other hand, a few users suggested that the usability must be improved for the app to be very useful for them. For example, a non-experienced software tester who has used the app daily provided the following comment "*Usability need (sic) to be improved*". Another experienced software developer noted the following: "*Poor usability, check standards designs from google*". Another user recommended that the app should include a "*single page summary*" to make the results more readable. We intend to follow these recommendations and improve the app's usability in the future. We discuss the Appsent's usability limitations in Section VII.

*RQ2. Does the newly developed app provide functionalities that are helpful?*

Our goal was to design an app that was effective in terms of the structure of the analytics that is provided, the data visualizations, the speed of the analytics process, and the cognitive load that was demanded by the end-users. For the structure of the analytics, we have observed that, in general, the evaluators found Appsent's analytics to be well-structured. The score returned was 3.4 (out of 4). Both experienced and non-experienced participants rated Appsent very similar for this dimension (experienced = 3.4, non-experienced = 3.3). For the cognitive load analysis, the average score was 3.4. Again, this score was nearly close to the maximum possible score of 4.0, which indicates that little cognitive load is required to understand and utilize Appsent. That said, we must also consider the video tutorial that was played before participants used the tool, as this may have enhanced evaluators' perception of Appsent. In addition, experienced evaluators scored Appsent much higher than those that lacked experience using analytics tools (experienced = 3.8, non-experienced = 2.8). This suggests that background knowledge of data analytics influenced evaluators' scores for Appsent positively. Since the experienced group had the understanding of the limitations of the data analysis, their expectations of Appsent's analytics results could have been less than the non-experienced group. Thus, reviewing helpful analytics features addressing the limitations of other tools may have resulted in the higher scores from this group.

Users were mostly happy with the *analytics* feature of the app. One user described the analytics part of the app as "*really good*" and "*provides fast summary allow to look at different levels*". Another user explains that the analytics is "*very useful*" and "*give(s) a clear understanding*". Another user mentioned that the app provides a "*very good analytic approach. Help(s) to react fast*" to users' reviews.

The data visualization score was found to be also high (3.5 out of 4). This score indicates that Appsent presented data visualization results adequately, which helped the participants to understand the opinions of end-users present in the app-reviews. Looking at the average score of the experienced and non-experienced groups of evaluators, we observed that the experienced group had a higher score (3.7) than the non-experienced group (3.2). This indicates that having prior knowledge of data visualization enabled the experienced group to understand Appsent's results better than the non-experienced group. The score for the speed of analytics was 3.4 out of 4. This score indicates that Appsent provides analytics results in a timely manner. Results were not very different across groups, although the experienced group returned a higher score (3.7) than non-experienced group (3.0). This suggests that, since the experienced group had prior knowledge about the complexity of text analytics, their expectations of the processing speed of Appsent were lower than the non-experienced group. By performing an overall analysis of the scores of the questionnaires that related to RQ2, Appsent application was regarded by the evaluators for its provision of helpful functionalities (Mean = 85.4, Median = 93.8, Standard Deviation of 17.1). Based on this overall score, and notwithstanding the prototypical nature of Appsent, we believe that analytics features of Appsent have been well designed and Appsent's data analytics features provided user-friendly support for the participants in understanding end-users' opinions present in the app-reviews.

*RQ3. Does the developed app provide suitable interpretative analytics on multiple levels?*

The research question aims to interpret the actual meaning behind the scores submitted by participants while answering the questionnaires that were related to data analytics features of Appsent (refer Table V). We specifically requested that participants evaluate the sentiment analysis, emotion analysis, timeline analysis, and co-occurring analysis features that are provided by Appsent. According to the evaluation results, the sentiment analysis outcomes provided by Appsent was found to be 3.1 out of 4. This score indicated that the sentiment analysis functionality helped the participants to understand the features that were well received by Appsent's end-users, and also those features that needed fixing. Evaluators that were experienced awarded a higher score (3.6) than the non-experienced group (2.5) for the sentiment analysis feature. A similar pattern was observed for the timeline analysis scores, where the experienced group graded Appsent higher (3.8) than the non-experienced group (3.0). These results indicate that the non-experienced participants had a higher expectation of Appsent sentiment analysis outcomes compared to experienced participants.

Participants' evaluation of Appsent's co-occurring features was 3.5 out of 4. From the group-based analysis, it was found that the average score of the co-occurrence analysis from the experienced group was relatively high (3.9 out of 4), compared to those of the non-experienced group, where the score was found to be 2.8 out of 4. This statistic for the experienced group was very close to the maximum score of 4. In fact, all participants stated that by knowing the defects/issues that were related to features helped them to gain a better understanding of end-users' opinions, and hence co-occurrence analysis in Appsent proved to be a strong data analytics feature. By analyzing the scores of all five questions that were aimed at answering RQ3, the average score was found to be 82 out of 100 (Median = 85.0 and Standard Deviation = 17.6). Based on this result, it may be inferred that Appsent application helped users to understand complex relationships between the entities of interest expressed by end-users. As highlighted in Section II, we have identified the gap where there was no co-occurrence analysis feature in the commercial applications available. This feature is innovative for any sentiment analysis app if users are to understand the relationships among end-users' opinions, features or issues [20]. Thus, we believe that co-occurrence analysis is a robust data analytics approach that can help to extract meaningful insights from app-reviews.

## VII. THREATS TO VALIDITY

We concede that there are limitations to the work that is presented here. This project strongly focused on data analytics, and thus Appsent was developed from that perspective as a proof of concept that a mobile app would be of utility as an analytics tool. To this end, we did not examine multiple interface designs before finally settling on a specific design. We acknowledge that the app (in it's current format) might have some usability limitations, as indicated by some users during the evaluation phase. However, improvement of Appsent's usability features is planned for the future releases. We also believe that there is scope to implement Appsent in a PC environment where there would be more processing power available and capacity for richer visualizations and overall enhanced usability. We are engaging colleagues at another New Zealand-based university with a view of implementing such an interface.

Only one set of app-reviews was analyzed as a part of this study, which may not provide a wide range of testing data for Appsent. However, as the data contained over 50,000 reviews, we believe that the data was adequate for the evaluations that were performed. Also, there were only fifteen participants present for the evaluation of Appent, which could potentially affect the external reliability of the study. We also accept that there is a possibility of the presence of miss-spelled features and issues that might have affected the result of the stemming operation in the text pre-processing stage. Lastly, the sentiment analysis and emotion analysis techniques used in this work may suffer from limitations, although these were previously assessed as adequate by numerous experiments (e.g., [6, 7]). On the likelihood that these techniques may not always perform accurately, there is a possibility that some of the features or issues could have been misclassified. This aspect need further consideration, and particularly in collaboration with our industry partners.

In this study, we have used a smartphone (LG Nexus 5) that has good hardware specifications (at the time that the study was conducted in May 2017). Hence, all the participants reported Appsent's data analytics and user interfacing performance to be fast, consistent and smooth. That said, there is scope to benchmark Appsent's performance on other devices with different hardware specifications.

## VIII. Conclusion

On the premise that there are numerous text analytics apps, but these lack pertinent features, we have designed and developed a prototype that potentially address several gaps in such apps. Our tool, Appsent, runs on mobile devices and analyses end-users' reviews logged about apps. We used text mining techniques and best practice design principles to provide a robust data analysis framework. By means of empirical evaluation, it was found that Appsent was successful in extracting knowledge of interest from app-reviews, and presented this knowledge in an understandable way through meaningful visualization methods. Appsent software may help providers (or developers) to quickly analyze users' feedback to make rapid decisions in terms of product improvements. Appsent is composed of well-structured modules, and generates efficient data analytics results. These results of the analytics can be visualized in meaningful ways. From these results, the product developers may determine which end-user requirements need urgent attention and fixing. A distinguishing feature of Appsent is its provision of functionalities that allow users to examine the relationships (co-occurring terms) between features and associated issues, which helps with understanding interconnections between features and issues. Based on the evidence provided in this paper, we believe that Appsent could be further developed to provide powerful analytics in understanding users' opinions in text reviews belonging to any domain, which could increase the response speed of product developers towards addressing end-users' requirements.